\documentclass[12pt]{article}

\renewcommand{\baselinestretch}{1.5}

\newcommand{\sth}{\sigma^3}
\newcommand{\stw}{\sigma^2}
\newcommand{\son}{\sigma^1}
\newcommand{\sfo}{\sigma^4}

\newcommand{\pmSig}{\,^{\pm}\!\Sigma}
\newcommand{\pOmega}{\,^+\!\Omega}

\newcommand{\pOmsthE}{\,^+\!\Omega_{\sigma^3{\rm E}}}
\newcommand{\mOmega}{\,^-\!\Omega}

\newcommand{\mOmsthE}{\,^-\!\Omega_{\sigma^3{\rm E}}}
\newcommand{\pmOmega}{\,^{\pm}\!\Omega}
\newcommand{\pmOmsth}{\,^{\pm}\!\Omega_{\sigma^3}}
\newcommand{\pmOmsthE}{\,^{\pm}\!\Omega_{\sigma^3{\rm E}}}
\newcommand{\pomega}{\,^+\!\omega}
\newcommand{\momega}{\,^-\!\omega}
\newcommand{\pmomega}{\,^{\pm}\!\omega}
\newcommand{\pmbvstw}{\,^{\pm}\!\mbox{\boldmath$v$}_{\sigma^2}}
\newcommand{\pmbvstwE}{\,^{\pm}\!\mbox{\boldmath$v$}_{\sigma^2{\rm E}}}
\newcommand{\pmbvE}{\,^{\pm}\!\mbox{\boldmath$v$}_{\rm E}}
\newcommand{\pmRstw}{\,^{\pm}\!R_{\sigma^2}}
\newcommand{\pmrstw}{\,^{\pm}\!r_{\sigma^2}}
\newcommand{\pbvstw}{\,^{+}\!\mbox{\boldmath$v$}_{\sigma^2}}
\newcommand{\pbvstwE}{\,^{+}\!\mbox{\boldmath$v$}_{\sigma^2{\rm E}}}
\newcommand{\pbtaustw}{\,^{+}\!\mbox{\boldmath$\tau$}_{\sigma^2}}
\newcommand{\mbtaustw}{\,^{-}\!\mbox{\boldmath$\tau$}_{\sigma^2}}
\newcommand{\pRstw}{\,^{+}\!R_{\sigma^2}}
\newcommand{\mbvstw}{\,^{-}\!\mbox{\boldmath$v$}_{\sigma^2}}
\newcommand{\mbvstwE}{\,^{-}\!\mbox{\boldmath$v$}_{\sigma^2{\rm E}}}

\newcommand{\mRstw}{\,^{-}\!R_{\sigma^2}}
\newcommand{\Rstw}{R_{\sigma^2}}
\newcommand{\taustw}{\tau_{\sigma^2}}
\newcommand{\Omsth}{\Omega_{\sigma^3}}
\newcommand{\OmT}{\Omega^{\rm T}}
\newcommand{\dotOm}{\dot{\Omega}}
\newcommand{\pmS}{\,^{\pm}\!S}
\newcommand{\pS}{\,^{+}\!S}
\newcommand{\mS}{\,^{-}\!S}

\newcommand{\sumstw}{\sum_{\sigma^2}}
\newcommand{\prodsth}{\prod_{\sigma^3}}

\renewcommand{\d}{{\rm d}}
\renewcommand{\Re}{{\mbox{Re\,}}}
\renewcommand{\Im}{{\mbox{Im\,}}}
\newcommand{\D}{{\cal D}}
\newcommand{\N}{{\cal N}}
\newcommand{\F}{{\cal F}}
\newcommand{\G}{{\cal G}}
\newcommand{\pN}{\,^{+}\!{\cal N}}
\newcommand{\mN}{\,^{-}\!{\cal N}}

\newcommand{\pmbphi}{\,^{\pm}\!\mbox{\boldmath$\phi$}}
\newcommand{\pmphi}{\,^{\pm}\!\phi}

\newcommand{\pmbv}{\,^{\pm}\!\mbox{\boldmath$v$}}
\newcommand{\pbv}{\,^{+}\!\mbox{\boldmath$v$}}
\newcommand{\mbv}{\,^{-}\!\mbox{\boldmath$v$}}
\newcommand{\bv}{\mbox{\boldmath$v$}}
\newcommand{\rv}{\mbox{\rm v}}
\newcommand{\rsv}{\mbox{\scriptsize\rm v}}

\newcommand{\bn}{\mbox{\boldmath$n$}}
\newcommand{\bsn}{\mbox{\scriptsize\boldmath$n$}}
\newcommand{\bl}{\mbox{\boldmath$l$}}
\newcommand{\bsl}{\mbox{\scriptsize\boldmath$l$}}
\newcommand{\br}{\mbox{\boldmath$r$}}
\newcommand{\bsr}{\mbox{\scriptsize\boldmath$r$}}
\newcommand{\bvarphi}{\mbox{\boldmath$\varphi$}}
\newcommand{\bpsi}{\mbox{\boldmath$\psi$}}

\newcommand{\pmv}{\,^{\pm}\!v}
\newcommand{\pv}{\,^{+}\!v}
\newcommand{\prvE}{\,^{+}\!{\rm v}_{\rm E}}
\newcommand{\mv}{\,^{-}\!v}
\newcommand{\mrvE}{\,^{-}\!{\rm v}_{\rm E}}
\newcommand{\gammaE}{\gamma_{\rm E}}
\newcommand{\pmrvE}{\,^{\pm}\!{\rm v}_{\rm E}}

\newcommand{\vstw}{v_{\sigma^2}}

\newcommand{\pmR}{\,^{\pm}\!R}
\newcommand{\pR}{\,^{+}\!R}
\newcommand{\mR}{\,^{-}\!R}

\newcommand{\dfun}{$\delta$-function }
\newcommand{\dfuns}{$\delta$-functions }
\newcommand{\desi}{\delta^6}
\newcommand{\deth}{\delta^3}

\newcommand{\de}{\delta}
\newcommand{\dejtw}{\delta^{(j+2)}}
\newcommand{\dej}{\delta^{(j)}}
\newcommand{\dejmodtw}{\delta^{(j({\rm mod}2)+2)}}
\newcommand{\dejmod}{\delta^{(j({\rm mod}2))}}

\newcommand{\sh}{{\rm sh}}
\newcommand{\ch}{{\rm ch}}

\newcommand{\laonaj}{l^{a_1}...l^{a_j}}
\newcommand{\naonaj}{n^{a_1}...n^{a_j}}
\newcommand{\daonaj}{\partial_{a_1}...\partial_{a_j}}

\newcommand{\ddg}{\frac{{\rm d}}{ {\rm d}g}}
\newcommand{\dxdg}{\frac{{\rm d} x}{ {\rm d}g}}
\newcommand{\brddgbr}{\left (\frac{{\rm d}}{ {\rm d}g} \right )}

\begin{document}

\title{Integration over connections in the discretized gravitational functional integrals}
\author{V.M. Khatsymovsky \\
 {\em Budker Institute of Nuclear Physics} \\ {\em
 Novosibirsk,
 630090,
 Russia}
\\ {\em E-mail address: khatsym@inp.nsk.su}}
\date{}
\maketitle
\begin{abstract}
The result of performing integrations over connection type variables in the path integral for the discrete field theory may be poorly defined in the case of non-compact gauge group with the Haar measure exponentially growing in some directions. This point is studied in the case of the discrete form of the first order formulation of the Einstein gravity theory. Here the result of interest can be defined as generalized function (of the rest of variables of the type of tetrad or elementary areas) i. e. a functional on a set of probe functions. To define this functional, we calculate its values on the products of components of the area tensors, the so-called moments. The resulting distribution (in fact, probability distribution) has singular ($\delta$-function-like) part with support in the nonphysical region of the complex plane of area tensors and regular part (usual function) which decays exponentially at large areas. As we discuss, this also provides suppression of large edge lengths which is important for internal consistency, if one asks whether gravity on short distances can be discrete. Some another features of the obtained probability distribution including occurrence of the local maxima at a number of the approximately equidistant values of area are also considered.

\end{abstract}




\newpage

\section{Introduction}

Suppose we are studying discrete version of a theory with (one of) the field variable being of the connection type. Example is Wilson formulation of the QCD\cite{Wils}. The discrete version of the connection is not an element of the Lee algebra but an element of the gauge group itself. In the path integral when integrating over the discrete connection type variable one is faced with integration over Haar measure on this group. In the case of (the first order formulation of) the Einstein gravity the group is that of the local rotational symmetry of the spacetime itself, i. e. SO(3,1). This is noncompact group while the Haar measure exponentially grows with Lorentz boost angles which together with Euclidean angles parameterize the discrete SO(3,1) connection. The corresponding integrations in the path integral should be made with care. The result of performing these generally does not exist as usual function (of the remaining variables of the tetrad type) but it can be given sense of the generalized function, or distribution (in fact, probability distribution for the remaining tetrad type variables). In fact, the latter turns out to include $\delta$-functions and their derivatives, but it also have a regular part being usual function. Such structure of this distribution does not allow to continue it to the Euclidean region ($\delta$-functions do not exist for complex arguments). This correlates with the fact that Euclidean path integral for pure Einstein gravity requires careful definition because of the unboundedness of the action from below.

Both discrete and continuum versions of the first order Einstein gravity are naturally formulated using decomposition of SO(3,1) element into two mutually complex conjugated elements of SU(2,C) or SO(3,C) (complex-valued). For definiteness, one might imply the following notations and sign conventions concerning splitting the tensors into (anti-)selfdual parts in the Minkowsky spacetime. Suppose there is SO(3,1) matrix (being connection type discrete variable),
\begin{equation}
\Omega = \exp {(\varphi^kE^a_{kb} + \psi^kL^a_{kb})}. 
\end{equation}
Its generator is expanded over the set of independent generators,
\begin{equation}
E_{kab} = -\epsilon_{kab}, ~~~ L_{kab} = g_{ka}g_{0b} - g_{0a}g_{kb}~~~(g_{ab} =
{\rm diag}(-1,1,1,1), \epsilon_{123} = +1 ). 
\end{equation}
We denote
\begin{equation}
\pmSig_{kab} = -\epsilon_{kab} \pm i(g_{ak}g_{0b} - g_{a0}g_{kb}) 
\end{equation}
so that
\begin{equation}
 ^*(\pmSig^{ab}) \! \equiv \frac{1}{2} \epsilon^{ab}_{~~cd}\pmSig^{cd} \! = \mp i\pmSig^{ab} ~ ( \epsilon^{0123} \! = \! +1), ~ \pmSig^a_{kb} \! \pmSig^b_{lc} \! = - \delta_{kl} \delta^a_c \! + \epsilon_{kl}^{~~m}
\pmSig^a_{mc}, 
\end{equation}
then
\begin{equation}
\Omega = \pOmega \mOmega, ~~~ \pmOmega = \exp \left (\frac{\varphi^k \mp i\psi^k} {2}\pmSig^a_{kb} \right ). 
\end{equation}
In the continuum limit the generator of $\Omega$ corresponds to $\omega^{ab}_{\lambda} \d x^{\lambda}$ where $\omega^{ab}_{\lambda} = - \omega^{ba}_{\lambda}$ is the infinitesimal connection which is thus expanded into self- and antiselfdual parts additively,
\begin{equation}
\omega^{ab} = \pomega^{ab} + \momega^{ab}, ~~~ ^*(\pmomega^{ab}) = \mp i\pmomega^{ab}, ~~~ \pmomega^{ab} = \frac{1}{2} \omega^{ab} \pm \frac{i}{4}
\epsilon^{ab}_{~~cd}\omega^{cd}. 
\end{equation}
For a triangle spanned by the two 4-vectors $l^c_1$, $l^d_2$ we can define {\it bivector} $v^{ab} = \frac{1}{2}\epsilon^{ab}_{~~cd}l^c_1l^d_2$. This variable (or antisymmetrized tetrad bilinears in the continuum theory) or used somewhere in the present paper its analytical continuation to arbitrary antisymmetric {\it area tensor} split in the same way,
\begin{equation}
v^{ab} = \pv^{ab} + \mv^{ab}, ~~~ \pmv^{ab} = \frac{1}{2} v^{ab} \pm \frac{i}{4}
\epsilon^{ab}_{~~cd}v^{cd}. 
\end{equation}
In particular,
\begin{equation}
2\pmv \circ \pmv = v \circ v \pm iv*v. 
\end{equation}
Here $A\circ B \equiv \frac{1}{ 2}A_{ab} B^{ab}$, $A*B$ $\equiv$ $\frac{1}{4}\epsilon_{abcd}A^{ab}B^{cd}$  for the two matrices $A, B$. The $^{\pm}$-parts map into three-dimensional vectors $ \pmbv $,
\begin{equation}\label{v-vec}
\pmv_{ab} \equiv \frac{1}{2}\pmv^k\pmSig_{kab}, ~~~ 2\pmv_k = - \epsilon_{klm}
v^{lm} \pm i(v_{k0} -v_{0k}).
\end{equation}
For a bivector $2 \pmbv = \pm i \bl_1 \times \bl_2 - \bl_1 l^0_2 + \bl_2 l^0_1$. Additional overall $i$ is here due to the fact that $v^{ab}$ is {\it dual} area tensor. Besides that,
\begin{equation}
\pmbv^2 = 2\pmv \circ \pmv. 
\end{equation}
The $\pmbv^2$ is $(-1)$ times the square of the (real for the spacelike triangle) area.

In the case of the continuum Einstein gravity the most general first order form of the action is that by Holst\cite{Holst,Fat},
\begin{eqnarray}\label{Holst}
\hspace{-5mm} S_{\rm Holst} & & = \frac{1}{8}\int{(\epsilon_{abcd}e^a_{\lambda}e^b_{\mu} \! + \frac{2}{\gamma}e_{\lambda c}e_{\mu d
})\epsilon^{\lambda\mu\nu\rho}
(\partial_{\nu}\omega_{\rho} \! - \partial_{\rho}\omega_{\nu} \! + \omega_{\nu}\omega_{\rho} \! - \omega_{\rho}\omega_{\nu})^{cd}{\rm d}^4x},
\end{eqnarray}

\noindent which generalizes the Cartan-Weyl form of the Einstein action and reduces to $\frac{1}{2} \int R\!$ $\!\sqrt{-g}\!$ $\!{\rm d}^4x$ in terms of metric $g_{\lambda\mu} = e^a_{\lambda} e_{a\mu}$ if the infinitesimal connection $\omega^{ab}_{\lambda}$ is expressed in terms of the tetrad $e^a_{\lambda}$ via equations of motion for $\omega^{ab}_{\lambda}$. Upon splitting $\omega$ into (anti-)selfdual parts the $S_{\rm Holst}$ splits too and is easily seen to have the form $(1 + i/\gamma) \pS_{\rm CW} + (1 - i/\gamma) \mS_{\rm CW}$ where $\pmS_{\rm CW}$ are (anti-)selfdual parts of the Cartan-Weyl continuum action  (that is, of $S_{\rm Holst}$ at $\gamma^{-1} = 0$), $\gamma$ is known as Barbero-Immirzi parameter\cite{Barb,Imm}.

The discrete minisuperspace formulation of general relativity on the piecewise flat manifolds or simplicial complexes is known as Regge calculus\cite{Regge,Cheeger}. Invoking the notion of discrete tetrad and connection first considered in Ref. \cite{Fro} we have suggested in Ref. \cite{Kha} representation of the Einstein action $\frac{1}{2} \int{R\sqrt{-g}{\rm d}^4x}$ on this manifold in terms of area tensors and finite rotation SO(4) (SO(3,1) in the Minkowsky case) matrices, and also in terms of (anti-)selfdual parts of finite rotation matrices. For the latter we write
\begin{equation}
\pmS = \sumstw \sqrt{\pmbvstw^2} \arcsin \frac{\pmbvstw * \pmRstw (\Omega )} {\sqrt{\pmbvstw^2} }.
\end{equation}

\noindent Here $\pmbvstw$ are area vectors (\ref{v-vec}) of the triangle $\stw$, in the Minkowsky case $\Omsth$ is connection SO(3,1) matrix on the tetrahedron $\sth$ which we call simply connection, $\Rstw$ is curvature matrix on the triangle $\stw$, holonomy of $\Omega$'s (some product of $\Omega$'s). For a 3-vector $\bv$ and a $3\times 3$ matrix $R$ we have denoted $\bv * R \equiv \frac{1}{2}v^a R^{bc} \epsilon_{abc}$, and for $\pmRstw$, the (anti-)selfdual part of $\Rstw$, we have used adjoint, SO(3) representation (to be precise, SO(3,C) matrix).

These representations result in the same $\frac{1}{2} \int{R\sqrt{-g}{\rm d}^4x}$ on the piecewise flat manifold in terms of the purely edge lengths upon excluding rotation matrices by classical equations of motion (that is, on-shell). Taking into account that in the Minkowsky case $\pS = ( \mS )^*$ we can write out the most general combination of $\pS$, $\mS$ which i) is real and ii) reduces to $\frac{1}{2} \int{R\sqrt{-g}{\rm d}^4x}$ in terms of the purely edge lengths on-shell, as $S$ = $C \pS + C^* \mS$ where $C + C^* = 1$, that is $C = [1 + i \cdot \mbox{(real parameter)}]/2$. We see that there is direct analogy with the discrete case if we write $C = (1 + i / \gamma)/2$ where the discrete analog of $\gamma$ is denoted by the same letter. We assume $0 < \gamma < \infty$. Thus
\begin{equation}\label{S+S}
S = \left (1 + \frac{i}{\gamma}\right )\frac{1}{2} \pS + \left (1 - \frac{i} {\gamma}\right )\frac{1}{2} \mS.
\end{equation}

Let us write out a discretized functional integral with this action, $\int \exp (iS) D q$, $q$ are field variables (some factors of the type of Jacobians could also be present). Functional integral approach in Regge calculus was earlier developed, see, e. g., Refs. \cite{Fro,HamWil1,HamWil2}. Suppose we have performed integration over rotation matrices and are interested in the dependence of the intermediate result on area tensors. Of course, different (components of) area tensors are not independent, but nothing prevent us from studying analytical properties in the extended region of varying these area tensors as if these were independent variables. Namely, consider integral
\begin{eqnarray}\label{intDOm}                                                     
\N = \int \exp \frac{i}{2} \sumstw \left [ \left (1 + \frac{i}{\gamma} \right ) \sqrt{\pbvstw^2} \arcsin \frac{\pbvstw * \pRstw (\Omega )}{\sqrt{\pbvstw^2} } \right. \nonumber\\ \left. + \left (1 - \frac{i}{\gamma} \right ) \sqrt{\mbvstw^2} \arcsin \frac{\mbvstw * \mRstw (\Omega )}{\sqrt{\mbvstw^2} } \right ] \prodsth \D \Omsth.
\end{eqnarray}

\noindent Matrices $\pmOmega$, $\pmR$ can be parameterized by complex vector angles $\pmbphi = \bvarphi \mp i\bpsi$ (rotation by the angle $\sqrt{\pmbphi^2}$ around the unit vector $\pmbphi /\sqrt{\pmbphi^2}$).

A special feature of the integrals like (\ref{intDOm}) is that these are generally not usual functions of $\pbv$, $\mbv = (\pbv)^*$ but rather generalized functions or distributions. This is because the Haar measure $\D \Omega$ grows exponentially with imaginary angles (i. e. on Lorentz boosts), and integral over it may diverge as usual function. This is consistent with the sense of the result of intermediate integrations in the path integrals over connections as some objects to be further integrated over area tensors. Therefore in what follows we just treat these integrals as distributions, i. e. study integrals of these expressions with some probe functions. For the probe functions taken as products of components of $\pmbv$ we can give sense to these integrals (called moments in this case). This allows to define distributions of interest. Remarkable is that the singular ($\delta$-function-like) part of the latter turns out to have support outside the physical region $\Im \bv^2 = 0$. So this distribution in the physical region is usual function.

Before proceeding with analysis of such integrals it is useful to maximally separate out the gauge degrees of freedom contained in the connection variables $\Omsth$. The matrices $\Rstw$ possess more physical sense. Only part of these are independent; those on certain set of the triangles $\F$ are functions of other $\Rstw$s, $\stw \not \in \F$ due to the Bianchi identities\cite{Regge}. A natural step is to reduce integration element $\prodsth \D \Omsth$ to $\prod_{\stw \not \in \F} \D \Rstw$ times the product of $\D \Omsth$'s on a certain set $\G$ of $\sth$s which absorb the rest of the gauge degrees of freedom, $\prod_{\sth \in \G} \D \Omsth$.

\begin{figure}[h]\unitlength 0.20mm
\begin{picture}(200,160)(-100,25)
\put (150,20){\line(0,1){120}} \put (150,20){\line(3,2){120}} \put
(150,20){\line(4,1){160}} \put (150,60){\line(1,1){120}} \put
(150,60){\line(2,1){160}} \put (150,60){\line(3,1){120}} \put
(150,60){\line(1,0){160}} \put (150,140){\line(3,1){120}} \put
(150,140){\line(1,0){160}} \put (270,100){\line(0,1){80}} \put
(270,100){\line(1,-1){40}} \put (270,180){\line(1,-1){40}} \put
(270,180){\line(1,-3){40}} \put (310,60){\line(0,1){80}} \put (310,135){$~l^{+}$} \put
(270,180){$~k^{+}$} \put (137,135){$i^{+}$} \put (142,55){$i$} \put (270,100){$~k$}
\put (310,55){$~l$} \put (132,15){$i^{-}$}
\end{picture}
\renewcommand{\baselinestretch}{1.0}
\caption{Fragment of the 3-prism.}\label{3prism}
\end{figure}
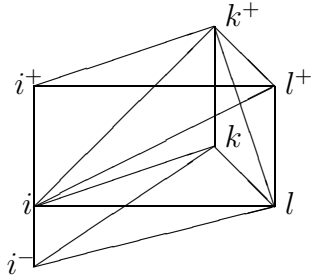

Let us illustrate this by explicitly expressing $\Omsth$s in terms of $\{ \Rstw : \stw \not\in {\cal F}\}$ and $\{ \Omsth : \sth \in \G \}$ for the simplicial structure regular in certain direction. This structure specifies the idea of constructing the piecewise flat 4-geometry of separate piecewise flat 3-geometries\cite{MisThoWhe}. The simplest such structure is widely used periodic simplicial decomposition when the spacetime is divided into 4-cubes and each 4-cube is divided into 24 4-simplices\cite{RocWil} by diagonals emitted from certain vertex of the cube. This structure is uniform w. r. t. the choice of vertex and similar along any of the four directions along the 4-cube axes. (If it is important to have finite number of vertices in lattice analysis, one might easily pass to the torus topology by imposing cyclic boundary conditions along corresponding directions.) We use our general notations\cite{Kha02} suitable for this structure as a particular case, see fig.\ref{3prism}.

Namely, we consider the sequence of $ t=const$ 3D
simplicial 3-geometries, call these {\it leaves}, separated by, say, the change $\Delta t$ of the coordinate $t$  such that for each vertex $ i$ in any of the
leaves there are both its image $ i^+$ and pre-image $ i^-$ in the two neighboring leaves at $t - \Delta t$ and at $t + \Delta t$. The coordinate $t$ is denoted as the (world) time although it well can be viewed as space coordinate. The points of the leaf at $t$ are denoted by the letters $i,~k,~l,\ldots$. Let us denote a simplex by enumerating its vertices in round brackets. Besides that, we
assume the same scheme of connection of different vertices by links
in the leaves. Let us call these links and other simplices like, e. g., tetrahedron $(iklm)$ completely contained in the leaf, the {\it leaf} simplices. The full 4D complex is supposed to result
from these 3D leaves by triangulating the space between each two
neighboring ones by new, {\it diagonal} and $t${\it -like} links. The link $(ii^+)$ or $(ii^-)$ which connects a vertex with
its image in the neighboring leaf, will be called $t${\it -like} link. Any other simplex like, e. g., the tetrahedron $(ikk^{+}l)$ containing such link, will be called $t${\it -like} simplex. The {\it diagonal} simplices are the remaining ones, e. g. tetrahedron $(ik^{+}lm)$.

The spacetime is thus divided into the "$t$-like" four-dimensional prisms. A face of the latter, the three-dimensional prism is just shown in fig.\ref{3prism}.

Then $\Rstw$ on the leaf and diagonal triangles $\stw$ could be taken as independent ones, and $\Rstw$ on the $t$-like triangles are functions of them via Bianchi identities. The $\Omsth$ on the leaf and diagonal tetrahedrons $\sth$ could be taken as purely gauge ones. (All this in particular implies that the number of the leaf and diagonal triangles matches the number of the $t$-like tetrahedrons. This is indeed so.) Indeed, consider a chain of expressions for the $\Rstw$ on the leaf and diagonal triangles $\stw$ sequently filling in a 3-prism, see fig.\ref{3prism},
\begin{eqnarray}\label{R-Omega}
& & \hspace{-18mm} \dots \dots \dots \dots \dots \dots \dots \dots \dots \dots \dots
\nonumber\\ R_{(ikl)} & = & \dots \OmT_{(i^-ikl)} \dots \Omega_{(ik^+kl)} \dots
\nonumber\\ R_{(ik^+l)} & = & \dots \OmT_{(ik^+kl)} \dots \Omega_{(ik^+l^+l)} \dots \\
R_{(ik^+l^+)} & = & \dots \OmT_{(ik^+l^+l)} \dots \Omega_{(i^+ik^+l^+)} \dots
\nonumber\\ & & \hspace{-18mm} \dots \dots \dots \dots \dots \dots \dots \dots \dots
\dots \dots \nonumber
\end{eqnarray}

\noindent The dots in the expressions for $R$ mean matrices $\Omega$ on the leaf and diagonal tetrahedrons. From Eqs. (\ref{R-Omega}) using some initial values for $\Omega$s on $t$-like tetrahedrons at $t = -\infty$ we can step-by-step express \dots
$\Omega_{(i^+ik^+l^+)}$ $\rightarrow$ $\Omega_{(ik^+l^+l)}$ $\rightarrow$
$\Omega_{(ik^+kl)}$ $\rightarrow$ $\Omega_{(i^-ikl)}$ $\rightarrow$ \dots (the arrow means "in terms of") as functions of $R$s and $\Omega$s on the leaf and diagonal triangles. In particular, knowing these $\Omega$s we can find
the curvatures on $t$-like triangles, the products of these $\Omega$s, e. g.
\begin{equation}                                                                   
R_{(i^+ik)} = \Omega^{\epsilon_{(ikl_n)l_{n-1}}}_{(i^+ikl_n)} \dots
\Omega^{\epsilon_{(ikl_1)l_n}}_{(i^+ikl_1)}.
\end{equation}

\noindent (This is just what shows up inside the 3D section as a curvature on the link $(ik)$.) Here $\epsilon_{(ikl)m}$ = $\pm 1$ is some sign function. Besides that, we can step-by-step rewrite integration element in another variables passing from the $\Omega$s on $t$-like tetrahedrons to $R$s on the leaf and diagonal triangles as \dots $\Omega_{(ik^+kl)}$ $\rightarrow$ $R_{(ikl)}$, $\Omega_{(ik^+l^+l)}$ $\rightarrow$ $R_{(ik^+l)}$, $\Omega_{(i^+ik^+l^+)}$ $\rightarrow$ $R_{(ik^+l^+)}$, \dots . On each step we have typical relation between certain matrices $R$ and $\Omega$ as $\Omega = \Gamma_1 R \Gamma_2$ where SO(3,1) matrices $\Gamma_1$ and $\Gamma_1$ are products of another matrices $\Omega^{\pm 1}$ and $R$ temporarily treated as constants. Then $\D \Omega = \D R$ due to the left-right invariance of the Haar measure. We thus arrive at \begin{equation}
\prodsth \D \Omsth = \prod_{\stw \not \in \F} \D \Rstw \prod_{\sth \in \G} \D \Omsth
\end{equation}

\noindent where $\F$ is the set of $t$-like triangles and $\G$ is the set of the leaf and diagonal tetrahedrons. Thus integration in (\ref{intDOm}) over connections could be replaced modulo integration over some connections taken as "gauge" ones by that one over independent curvatures.

The present paper simply calculates the simplest integral of the type of Eq. (\ref{intDOm}). Outside of nonphysical singularities this function exponentially decays at large areas. 
Some features of this function (Euclidean version, occurrence of a number of local maxima), consequences for the general case, for the measure on the edge lengths,  are discussed.

\section{The moments of probability distribution}\label{moments}

\subsection{Defining the moments}

As mentioned above, we consider Eq. (\ref{intDOm}) as function of arbitrary $\pbv$, $\mbv = (\pbv)^*$. Consider moments of $\N$ (integral with powers of $\pmbv$). As also mentioned, $\N$ contains $\delta$-functions which cannot be continued to/from the complex arguments. However, in the definition of moments the contours of integrations can be deformed to complex plane. Namely, integration contour over $\bpsi$ considered as complex variable is deformed from real to imaginary values. The $\pmbphi$ become independent real variables, and integration over $\D \Omega$ splits,
\begin{equation}
\D \Omega = \D \pOmega \D \mOmega, ~~~ \N = \pN \mN, ~~~ \D \pmOmega = \frac{\sin^2 (\pmphi/2)}{ 4\pi^2 \pmphi^2} \d^3 \pmbphi.
\end{equation}

\noindent We define $\N$ in the region where $\bl$ = $(1 \pm i/\gamma)\pmbv /2$ are real and then continue to $\pmbv$ of interest. We are faced with integrals of the form
\begin{equation}\label{M}
\int \D \pmR ... \int e^{ilg(\bsn\bsr)} \laonaj \d^3 \bl
\end{equation}

\noindent where $l = \sqrt{\bl^2}$, $\bn = \bl / l$, $g(x)$ is odd analytical in the neighborhood of $x = 0$ function, $g(-x) = -g(x)$, such as principal value of $\arcsin x$ of interest or simply $x$. The vector $r^a = \epsilon^a{}_{bc} \pmR^{bc} / 2 = (\pmphi^a \sin \pmphi) / \pmphi$. Existence of integrals (\ref{M}) easily follows at $g(x) = x$: we simply get derivatives of \dfuns $\deth (\br )$ which are then integrated over
\begin{equation}
\D \pmR = \left (\frac{1}{ \sqrt{1 - r^2}} - 1 \right ) \frac{\d^3 \br }{ 8\pi^2 r^2}.
\end{equation}

\noindent Finiteness is provided by analyticity of this measure at $r \to 0$, $\D \pmR = (c_0 + c_1 r^2 + c_2 (r^2)^2 + ...) \d^3 \br$. Dots in $\int \D R ...$ in (\ref{M}) mean possible dependence on $R$ of factors provided by $\Rstw$ on other triangles $\stw$ due to the Bianchi identities.

For the more general case $g(x) \neq x$ (\ref{M}) is also finite. Again, consideration may go through appearance of \dfuns at an intermediate stage. Namely, special structure of exponential (\ref{M}) allows to extend integration over $l$ to the whole range $( -\infty , +\infty )$. This is only possible because formal substituting $l \to -l$ is equivalent to $\bn \to -\bn$ due to the oddness of $g(x)$. This results in \dfuns of $g$,
\begin{eqnarray}\label{intlg}
 & & \int e^{ilg(\bsn\bsr)} \laonaj \d^3 \bl = \frac{1 }{ 2} \int \naonaj \d^2 \bn \int\limits^{+\infty}_{-\infty} e^{ilg(\bsn\bsr)} l^{j+2} \d l \nonumber\\ & &  \phantom{\int e^{ilg(\bsn\bsr)} \laonaj \d^3 \bl} = \frac{1 }{ 2} (2\pi) (-i)^{j+2} \int \dejtw (g(\bn \br)) \naonaj \d^2 \bn.
\end{eqnarray}

\noindent Apply $\dejtw (g(x))$ to probe functions,
\begin{equation}\label{probe}
\int \dejtw (g(x)) f(x) \d x = \left ( -\ddg \right )^{j+2} \left [ f(x(g)) \dxdg \right ]_{g=0},
\end{equation}

\noindent we find it be expanded over $\dejtw (x)$, $\dej (x)$, ... , $\dejmodtw (x)$, $\dejmod (x)$. For each term in this expansion in the RHS of the formula (\ref{intlg}) we can use analogous to Eq. (\ref{intlg}) relation (at $g(x) = x$) but read from right to left for the backward conversion to the expression having form of the Fourier transform of power function of $\bl$, $l$. The leading term $\dejtw (\bn \br )$ in this expansion is converted to $\int \exp (i \bl \br )\!$ $\!\laonaj\!$ $\!\d^3 \bl\!$ $\!= (2\pi)^3 (-i)^j \daonaj \deth (\br )$ (the LHS of the Eq. (\ref{intlg}) at $g(x) = x$). This being integrated over $\D \pmR$ is finite. Subsequent terms $\de^{(j-2k)} (\bn \br )$, $0 \leq k \leq  [j/2]$ ([j/2] is integer part) are converted to $(-i)^{j+2} \daonaj 2\pi^2\!$ $\!(r^2\!$ $\!+ \varepsilon^2)^{k-1/2} [(2k)!]^{-1}$. Here intermediate regularization $\varepsilon \to 0$ of the singularity at $\br = 0$ is convenient to detect possible formation of the contact terms via $\partial^2_a (1/r)$ = $-4\pi \deth (\br )$. Finiteness is provided by analyticity of the measure at $r = 0$,
\begin{equation}
\D \pmR = (P_n (r) + f_n (r)) \theta (1-r) \d^3 \br, ~~~ P_n = \sum^n_{m=0} c_m r^{2m}, ~~~ f_n = O(r^{2n+2}),
\end{equation}

\noindent $\theta (y)$ is Heaviside step function. Convergence of integral with $f_n (r) \theta (1-r)$, $n = [j/2] - k$, is seen immediately, and that of integral with $P_n (r) \theta (1-r)$ follows upon multiple applying integration by parts.

\subsection{The moments of the simplest integral over connections}

Above we have shown possibility to define the (nonabsolutely convergent) expressions for the moments in a finite way. If simplified integral is considered,
\begin{equation}\label{N_0}
\N_0 = \int \exp \frac{i}{ 2} \left [ \left ( 1 \! + \! \frac{i }{ \gamma } \right ) \sqrt{\pbv^2} \arcsin \frac{ \pbv \! * \! \pR }{ \sqrt{\pbv^2}} \! + \! \left ( 1 \! - \! \frac{i }{ \gamma } \right ) \sqrt{\mbv^2} \arcsin \frac{ \mbv \! * \! \mR }{ \sqrt{\mbv^2}} \right ] \D R
\end{equation}

\noindent (a dependence on $R$ of $\Rstw$ on other triangles $\stw$ due to the Bianchi identities is not taken into account), calculation of arbitrary moment of it can be performed in closed form. It is sufficient to integrate in (\ref{M}) and then over $\D \pmR$ the scalars $l^{2k}$. Substituting $f(x) = f^{(2m)} (0) x^{2m} / (2m)!$ and $j = 2k$ into (\ref{probe}) gives $(\d / \d g)^{2k+3} [x^{2m+1} (g)] / (2m+1)! |_{g=0}$ for the coefficients in the expansion of $\de^{(2k+2)} (g(x))$ over $\de^{(2m)} (x)$s. Using (\ref{intlg}) read from right to left we reduce $\int \de^{(2m)} (\bn \br ) \d^2 \bn$ to $\int \exp (i \bl \br ) l^{2m - 2} \d^3 \bl$. Summation over $m$ with the coefficients found allows to find $\int \exp [i l g(\bn \br )] l^{2k} \d^3 \bl$ so that
\begin{eqnarray}\label{moment}
 & & \int \D \pmR \int e^{ilg(\bsn\bsr)} l^{2k} \d^3 \bl \nonumber\\ & & = (-1)^{k+1} \int \D \pmR \int e^{i\bsl\bsr} \d^3 \bl \brddgbr^{2k+3} \left [ \sum^{\infty}_{m=0} (-1)^m \frac{x^{2m+1} (g)}{ (2m+1!)} l^{2m-2}\right ]_{g=0} \nonumber\\ & & = (-1)^{k+1} \brddgbr^{2k+2} \left [ \dxdg I(x)\right ]_{g=0}
\end{eqnarray}

\noindent with the "generating function"
\begin{eqnarray}\label{I}
 & & I(x) = \int \left (\frac{1}{ \sqrt{1 - r^2}} - 1 \right ) \frac{\d^3 \br }{ 8\pi^2 r^2} \int e^{i\bsl\bsr} \frac{\cos (xl) }{ l^2} \d^3 \bl \nonumber\\ & & = \pi \int\limits^1_x \left (\frac{1}{ \sqrt{1 - r^2}} - 1 \right ) \frac{\d r }{ r} = \pi \ln (1 + \sqrt{1 - x^2}).
\end{eqnarray}

\noindent We have extended summation to infinite number of powers of $x$ keeping in mind that upon applying $(\d / \d g)^{2k+3} (\cdot )_{g=0}$ only finite number of terms are active.

\section{Restoring probability distribution from the moments}\label{restoring}

In the simplest case $x = g$ the moment (\ref{moment}) is
\begin{equation}
\pi (-1)^{k+1} \brddgbr^{2k+1} \left (\frac{1 }{ g} - \frac{1 }{ g \sqrt{1 - g^2}}\right )_{g=0}.
\end{equation}

\noindent At $x = \sin g$ the moment is
\begin{equation}
\pi (-1)^{k+1} \brddgbr^{2k+2} \left [ \cos g \ln \left (1 + \cos g\right )\right ]_{g=0}.
\end{equation}

\noindent It is not difficult to find out density of distribution giving these values on monomials (perform a kind of Mellin transform). Appropriate entries in the table of integrals are
\begin{equation}
\frac{1 }{ \sqrt{1-g^2}} = \frac{2}{ \pi} \int\limits^{\infty}_0 \ch gl K_0(l) \d l,
\end{equation}

\noindent $K_0$ is modified Bessel function, and
\begin{equation}
\frac{g}{ 2} \sin g - \frac{1}{ 2} + \frac{1}{ 2} \cos g \ln [2(1 + \cos g)] = \int\limits^{\infty}_0 \frac{l }{ l^2 + 1} \frac{\ch gl}{ \sh \pi l} \d l.
\end{equation}

\noindent Thus we find at $g(x) = x$
\begin{equation}\label{my3d94}
\int \D \pmR \int e^{i\bsl \bsr} f ( l^2 ) \d^3 \bl \propto \int \frac{Ki_1(l)}{ l}f(-l^2) \frac{\d^3 \bl}{ \pi^2}
\end{equation}

\noindent where
\begin{equation}
Ki_1(l) = \int\limits^{\infty}_0 e^{-l\ch \eta} \frac{\d \eta}{ \ch \eta} = \int\limits^{\pi /2}_0 \exp \left (-\frac{l}{ \sin \varphi}\right ) \d \varphi
\end{equation}

\noindent is integral of $K_0(l) = -[Ki_1(l)]^{\prime}, Ki_1(\infty ) = 0$. This had appeared in our work\cite{Kha2} in 3D SO(3) Regge calculus.

At $g(x) = \arcsin x$ of present interest
\begin{equation}
\int \D \pmR \int e^{ilg(\bsn \bsr)} f ( l^2 ) \d^3 \bl \propto \int \frac{\pi l }{ \sh \pi l} \frac{f(-l^2) }{ l^2 + 1} \frac{\d^3 \bl }{ \pi^2} + (2 \ln 2 - 4) f(1) - 4f^{\prime} (1).
\end{equation}

\noindent The RHS in both cases is normalized so that appearing there integral be 1 at $f$ = 1. For the integral (\ref{N_0}) we deform integration contour and pass from $\bl$ to $\bv \equiv \pbv = 2\bl (1 + i/\gamma )^{-1}$ and $\bv^*$,
\begin{equation}
\int \N_0 (\rv, \rv^*) f(\rv^2)h(\rv^2)^*\d^3 \bv \d^3 \bv^* = \mu (f) \mu (h)^*,
\end{equation}

\noindent where $\rv = \sqrt{\bv^2}$ ~ ($\rv^2 + \rv^{*2} = -{\rm Tr} (v^2)$) and
\begin{eqnarray}\label{mu-f}
 & & \mu (f) = \frac{i}{ 2} \int \frac{(1/\gamma - i) \rv/2}{ (1/\gamma - i)^2 \rv^2/4 + 1} \frac{f(\rv^2) \d^3 \bv }{ \sh [\pi (1/\gamma - i) \rv/2]} \nonumber\\ & & \hspace{-1cm} + 4\pi (1 + i/\gamma)^{-3} [(\ln 2 - 2) f(4(1 + i/\gamma)^{-2}) - 8(1 + i/\gamma)^{-2}f^{\prime}(4(1 + i/\gamma)^{-2})].
\end{eqnarray}

\noindent Here additional to integral terms have support at nonphysical points $\rv^2 = 4(1 + i/\gamma)^{-2}$. Thus in any region excluding these points (e. g., in physical region $\Im \rv^2 = 0$) we have
\begin{equation}\label{Nvv}
\N_0 (\rv, \rv^*) = \left | \frac{1 }{ \frac{1 }{ 4}\left (\frac{1}{ \gamma} - i\right )^2 \rv^2 + 1} \cdot \frac{\frac{1 }{ 4}\left (\frac{1}{ \gamma} - i\right ) \rv }{ \sh \left [\frac{\pi }{ 2} \left (\frac{1}{ \gamma} - i\right ) \rv \right ]} \right |^2.
\end{equation}

Looking at the logarithmic expression for generating function (\ref{I}), it is interesting to ask to what extent our definition based on continuation from Euclidean-like region relies on the compactness of SO(4). Alternative way to define the moments of (\ref{N_0}) might proceed via deforming integration contours so that $\bvarphi$ be imaginary. The $\pmbphi$ become independent {\it imaginary} variables (pure Lorentz boost angles), and integration over $\D R$ splits. The $\br$ is imaginary, we define moments on imaginary $\bl$ and find $I(x)$ at imaginary $x$. Logarithmic divergence at $r \to \infty$ appears. Let the cut off for $r$ in (\ref{I}) after replacing $r \to ir$ be $r_0$. Then continue $I(x)$ to real $x$. The $I(x)$ (\ref{I}) is modified by adding constant to logarithm,
\begin{equation}
I(x) \Rightarrow \pi \ln \left (\frac{1 + \sqrt{1 - x^2} }{ 1 + \sqrt{1 + r_0^2}}\right ).
\end{equation}

\noindent However, it is not difficult to see that this constant only modifies the coefficient at $f(4(1 + i/\gamma)^{-2})$ in (\ref{mu-f}) and thus does not affect the result (\ref{Nvv}).

In physical spacelike region $\rv^2 = \rv^{*2} = -|\rv|^2$, $|\rv|$ is module of the triangle area, the Eq. (\ref{Nvv}) behaves as $\exp (-\pi |\rv|)$ at large $|\rv|$. In physical timelike region $\rv^2 = \rv^{*2} = |\rv|^2$, this behaves as $\exp (-\pi |\rv|/\gamma)$ at large $|\rv|$. For comparison, replacing $\arcsin x$ by $x$ as in (\ref{my3d94}) gives $\exp (- |\rv|)$ and $\exp (- |\rv|/\gamma)$, respectively.

An alternative way to define moments does not include transformation to the Euclidean signature\cite{Kha22}. This approach encounters more singularities and requires careful mathematical analysis. The resulting probability distribution coincides with the present one.

\section{Discussion}

\subsection{Qualitative explanation of exponential suppression}

The above exponential suppression over areas arises already in the model integral
\begin{equation}
\int\limits^{+\infty}_{-\infty} e^{i\sqrt{-\rsv^2} \sh \psi}\d \psi = 2K_0(\sqrt{-\rv^2}).
\end{equation}

\noindent Here $\sqrt{-\rv^2}$ is modeling module of the spacelike area, $\psi$ is modeling Lorentz boost angle. This behaves as $\exp (-\sqrt{-\rv^2})$ at large $\rv^2$. Nonzero $\gamma^{-1}$ mixes spacelike and timelike area components and leads to exponential suppression also in the timelike region. Taking integrals over connections (\ref{intDOm}) also reminds calculating Fourier transform of 1, although on curvy (group) manifold. While usual Fourier transform gives $\delta$-function, limiting case of exremely rapidly decreasing exponent, calculations on curvy manifold lead to broadening this $\delta$-function. This is qualitative "physical" explanation of the exponential suppression for areas. In particular, when $g(x) = \arcsin x$, rotation angles enter exponential in the form close to the angles themselves. Then connection integral is more close to the usual Fourier transform and provides larger exponential suppression for areas than the integral at $g(x) = x$ (when exponential contains angles in the more nonlinear manner, as hyperbolic or trigonometric functions).

\subsection{Euclidean version}

Our definition of the integration over connection suggests that the Feynman path integral itself is defined (on area tensor monomials, in a theory where these are treated as independent ones). However, direct evaluation of the Euclidean path integral fails because of the unboundedness of the action, although there are indications that the problem can be handled in the lattice theory, via appearance of a non-trivial ultraviolet fixed point (see, e. g., review \cite{Ham}). We could modify definition. Namely, the Euclidean version of the path integral with $\exp (-S_{\rm E})$, with the Euclidean signature of space-time and SO(4) connection could be defined if integration contours are deformed via formal change of the integration variables $v \to -iv$\cite{Kha3}. Namely, supplying (real) Euclidean variables with subscript E, $S_{\rm E}$ follows formally from $S$ (\ref{S+S}) appearing in (\ref{intDOm}) by substitution $\pmbvstw \to \pmbvstwE$, $\pmOmsth \to \pmOmsthE$, $\gamma \to i \gamma_{\rm E}$ (the latter because the coefficients $(1 \pm i/\gamma)$ should become real). Now 3-vectors $\pbvstwE$ and $\mbvstwE$ are {\it independent} variables, as well as the elements of real SO(3) $\pOmsthE$ and $\mOmsthE$ are. The appropriate path integral can be defined on the area tensor monomials by deforming integration contours to purely imaginary tensors, $\pmbvstwE \to -i\pmbvstwE$ so that the monotonic exponent $\exp (-S_{\rm E})$ becomes oscillating one. After finding $\N$ or, more exactly, $\N_{\rm E}$ in this region, we should return to the original region of tensor values, $\pmbvstwE \to i\pmbvstwE$. Along the lines of sections \ref{moments}, \ref{restoring} thus defined Euclidean version of Eq. (\ref{Nvv}) reads
\begin{eqnarray}\label{NvvE}
& & \N_{0{\rm E}} (\prvE, \mrvE) = \\ & & \hspace{-5mm} \frac{\frac{1 }{ 4}\left (\frac{1}{ \gammaE} \! + \! 1\right ) \! \prvE }{\left [\frac{1 }{ 4}\left (\frac{1}{ \gammaE} \! + \! 1\right )^2 \!\!\!\! \prvE^2 \! + \! 1 \right ] \sh \left [\frac{\pi }{ 2} \left (\frac{1}{ \gammaE} \! + \! 1\right ) \!\! \prvE \right ]} \cdot \frac{\frac{1 }{ 4}\left (\frac{1}{ \gammaE} \! - \! 1\right ) \mrvE }{ \left [ \frac{1 }{ 4}\left (\frac{1}{ \gammaE} \! - \! 1\right )^2 \!\!\!\! \mrvE^2 \! + \! 1 \right ] \sh \left [\frac{\pi }{ 2} \left (\frac{1}{ \gammaE} \! - \! 1\right ) \!\! \mrvE \right ]}. \nonumber
\end{eqnarray}

\noindent Here $\pmrvE = \sqrt{\pmbvE^2}$. In physical region ($v^{ab}_{\rm E}$ is {\it bivector}) $\prvE = \mrvE \equiv {\rm v}_{\rm E}$. Using here Minkowsky $\gamma$, $\gammaE \to -i\gamma$, we reproduce suppression $\exp (-\pi {\rm v}_{\rm E})$ of the triangle with area ${\rm v}_{\rm E}$ in the {\it spacelike} region, as it should be for the Euclidean version. To continue Euclidean result to the {\it timelike} region, one should know it analytically beyond the physical region $\prvE = \mrvE$.


\subsection{General case}\label{GenCase}

In general, we have suppression of the triangles of large areas at least for part of the triangles, such that the resting part might be taken as those passing through the links being discrete analogs of lapse-shift.

In more detail, when returning from the simplest integral (\ref{N_0}) to the general one (\ref{intDOm}) the moments (\ref{moment}) get naturally generalized to the integrals of $\N$ with monomials of the components $\vstw^{ab}$ for the whole set of the 4-simplices $\stw$. Important new point is that this integration should not be over {\it all} $\d^6 \vstw$s. This is especially clearly seen in the case of the simplified action with the function $g(x) = \arcsin x$ replaced by $g(x) = x$. Then integrating exponent in (\ref{intDOm}) over $\d^6 \vstw$ with the components $\vstw^{ab}$ for a given $\stw$ leads to the derivatives of the \dfun $\desi (\Rstw - \Rstw^{\rm T})$. The set of holonomies $\{ \Rstw : \stw \supset \son \}$ for a given link $\son$ obeys Bianchi identities. Integral over $\prod_{\stw \supset \son} \d^6 \vstw$ would result in the singularity of the type of $[\desi (\Rstw - \Rstw^{\rm T})]^2$ for some $\stw \supset \son$. To avoid this, we should omit integration over any one of the triangle containing the link $\son$. In overall, well-defined integral of $\N$ times monomial of all $\vstw^{ab}$s is that over area tensors of the triangles $\stw$ not contained in some set $\F$, $\prod_{\stw \not\in \F} \d^6 \vstw$. It is not difficult to see that this is the same set $\F$ of the triangles considered in the Introduction on which curvatures could be expressed in terms of the curvatures on other triangles. Generally it can be defined as the set of the triangles with edges constituting a set of non-intersecting and non-self-intersecting unclosed broken lines (that is, only two such edges can meet at any vertex) passing through all the vertices of our manifold. We can refer for a moment to this set of links as to the discrete analogs of the lapse-shift vectors which pick up direction of a coordinate $t$ at each vertex. So the triangles $\F$ might be called "$t$-like" ones. Omitting integration over some area tensors is equivalent to saying that these are fixed. This resembles the issue of the gauge fixing in the continuum theory. There is even direct relation to such gauge fixing. Indeed, we can pass to the continuous time theory by shrinking the edges along any direction temporarily chosen as time. In this limit we can develop Hamiltonian formalism and find\cite{Kha3} that fixing area tensors of the triangles of $\F$ is an admissible gauge fixing in the resulting theory. This is analogous to fixing lapse-shift vectors in the continuum general relativity with additional difference that now when studying analytical properties of $\N$, as mentioned in Introduction, area tensors are treated as formally independent variables generally not corresponding to certain lapse-shift vectors, therefore fixed are namely area tensors of the $t$-like triangles, not lapse-shift vectors.

Thus, the moments of the distribution $\N$ can be defined as integrals of $\N$ with area tensor monomials over $\d^6 \vstw$, $\stw \not\in \F$. We have seen this for a simplified form of action with $g(x) = x$. At $g(x) = \arcsin x$ restrictions on the set of area tensor components to integrate over may be relaxed (since instead of $\desi (\Rstw - \Rstw^{\rm T})$ we have less singular distribution when integrating exponent over $\d^6 \vstw$), and this set might be larger than $\{ \vstw : \stw \not\in {\cal F}\}$; this point requires further studying. In any case, omitting integrations over $\d^6 \vstw$, $\stw \in \F$ is sufficient to avoid singularity when integrating exponent over $\d^6 \vstw$s. Next, it is more illustrative to redenote area tensors of the triangles of $\F$ as $\vstw \to \taustw$ to stress that these are temporarily referred to as constants. Contribution of the triangles of $\F$ in $\exp (iS)$ decouples multiplicatively as
\begin{eqnarray}\label{exp-tau}
& & \exp \frac{i}{2} \sum_{\stw \in \F} \left [ \left (1 + \frac{i}{\gamma} \right ) \sqrt{\pbtaustw^2} \arcsin \frac{\pbtaustw * \pRstw (\{ \Rstw : \stw \not\in {\cal F}\} , \{ \Omsth : \sth \in \G \} )}{\sqrt{\pbtaustw^2} } \right. \nonumber\\ & & \hspace{-2mm} \left. + \left (1 \! - \! \frac{i}{\gamma} \right ) \! \sqrt{\!\! \mbtaustw^2} \arcsin \! \frac{\! \mbtaustw \! * \!\! \mRstw (\{ \Rstw : \stw \not\in {\cal F}\} , \! \{ \Omsth : \sth \in \G \} )}{\sqrt{\mbtaustw^2} } \! \right ] \prodsth \D \Omsth
\end{eqnarray}

\noindent where $\Rstw$, $\stw \in \F$, is written by Bianchi identities as function of other $\Rstw$s, $\stw \not\in \F$ and $\Omsth$s, $\sth \in \G$. The earlier neglected dots in (\ref{M}) are just due to (\ref{exp-tau}). This is analytical function of $\pmrstw^a = \epsilon^a{}_{bc} \pmRstw^{bc} / 2$, analytical at $\pmrstw^a = 0$, $\stw \not\in \F$.

Finiteness of the integral of a function with any product of its arguments means that this function is decreasing faster than any inverse power of arguments. The simplest such function is exponentially decreasing one. This type of decreasing is most natural in the present case when the function of interest $\N$ is itself integral of exponent (and, by proper deformation of integration contours in complex plane, is expected to be representable as a priori combination of increasing and decreasing monotonic exponents). Exponential suppression should take place at least over area tensors of a part of triangles when area tensors of other triangles $\F$ (for which we can take $t$-like ones for any coordinate $t$) are fixed. In the particular case $\taustw = 0$, if $\stw \in \F$, we simply get factorization of $\N$ into the above calculated exponentially decaying factors for separate triangles $\stw \not \in \F$.

\subsection{Suppression of the edge lengths}

Examining the measure over link lengths we find suppression factors for degenerate (zero volume) configurations of spacetime, and, on the other hand, the number of triangles suppressed at large areas at present dimensionality four of spacetime turns out to be large enough to provide also suppression of large edge lengths.

In more detail, consider area tensor part of the path integral measure. The sets of variables $\{ \vstw : \stw \not\in {\cal F}\}$ and $\{ \Omsth : \sth \not\in \G \}$ become true canonically conjugate sets in the limit when some coordinate (not necessarily $t$) is made continuous (by shrinking simplicial sizes along it) and regarded as time in the Hamiltonian analysis\cite{Kha3}. We are trying to define the full discrete measure which would become true canonical one of the type of $\d p \d q$ in the continuous time limit. As usual for discrete counterpart of a continuum theory, some first class (commuting) constraints convert into non-commuting second class ones which, however, now remain first class in the flat spacetime. Then the constructed in the standard way canonical measure is singular at the flat spacetime. The way round this is to construct the measure in the theory with independent area tensors which, accidentally, have proved to be good also for defining integrals over connections in the present paper. As a result, we get a theory with first class constraints (at least, in empty spacetime). The constructed measure is simple and has the area tensor part $\prod_{\stw \not\in \F} \d^6 \vstw$. Then the measure could be projected onto the hypersurface of genuine gravity in the configuration superspace singled out by geometric relations between area tensors required for that the edge vectors would exist on which area tensors could be constructed as bivectors.

To illustrate this in the notations somewhat close to the continuum ones using tetrad, we again assume certain ordering of the simplicial structure w.r.t. a coordinate $t$ direction, see fig.\ref{3prism}. Then each 4-simplex has a $t$-like edge (i. e. of the type of $(i^+ i)$). A 4-simplex is spanned by a tetrad $l^a_{\lambda}$ in the 4-simplex local frame emanating from some of its vertices. Let $l^a_0$ be 4-vector of the $t$-like edge, an analog of the lapse-shift vector. The bivectors are $v^{ab}_{\lambda\mu} \equiv \frac{1}{2} \epsilon^{ab}{}_{cd} l^c_{\lambda} l^d_{\mu}$. The triangles having $v^{ab}_{0 \alpha}, \alpha = 1, 2, 3$ and analogous in other 4-simplices form the set $\F$. The area tensors $v^{ab}_{\alpha} \equiv \frac{1}{2} \epsilon_{\alpha}{}^{\beta \gamma} v^{ab}_{\beta \gamma}, \alpha, \beta, \gamma = 1, 2, 3$ are those over which integration element
\begin{equation}\label{d6v}
\prod_{\alpha} \d^6 v^{ab}_{\alpha} \sim \prod_{\alpha} \d^3 \bv_{\alpha} \d^3 \bv^*_{\alpha}
\end{equation}

\noindent enters the measure. Here $\d^3 \bv \d^3 \bv^* \equiv 2^3 \d^3 \Re \bv \d^3 \Im \bv$. To confine ourselves by only those $v^{ab}_{\alpha}$ which are indeed bivectors for certain tetrad, we insert $\delta$-function factor $\desi (\bv_{\alpha} \cdot \bv_{\beta} - \bv^*_{\alpha} \cdot \bv^*_{\beta}) (\det \| \bv_{\alpha} \cdot \bv_{\beta} \|)^2$. This simply enforces the conditions $\epsilon_{abcd} v^{ab}_{\alpha} v^{cd}_{\beta} = 0$ ensuring existence of a triad $l^a_{\alpha}$ defining $v^{ab}_{\alpha}$. We imply in the standard way the Shwinger "time gauge"\cite{Schw} $l^0_{\alpha} = 0, \alpha = 1, 2, 3$ so that the solution is $2 \bv_1 = i \bl_2 \times \bl_3, \dots, {\rm perm} (1, 2, 3), \dots$. The multiplier of $\delta$-function is singled out by symmetry consideration that the overall $\delta$-factor be invariant w.r.t. an arbitrary redefinition of the triad $\bv_{\alpha} \to A_{\alpha}{}^{\beta} \bv_{\beta}$ leaving three-dimensional section in the same hyperplane, i. e. with real $A_{\alpha}{}^{\beta}$. With this factor, the integration element reduces to
\begin{equation}\label{d3v-det}
|\det \| \bv_{\alpha} \cdot \bv_{\beta} \| |^{3/2} \prod_{\alpha} \d^3 \bv_{\alpha}
\end{equation}

\noindent (up to integration over 3 rotational degrees of freedom of the triad $\bv^*_{\alpha}$).

Note that the measure is often considered as definable up to some power of $-\!$ $\!\det\!$ $\! \| g_{\lambda\mu} \|$. Now $g_{\lambda\mu} = l^a_{\lambda} l_{a \mu}$ in the 4-simplex in certain piecewise-affine coordinates $x^{\lambda}$. A positive power of $-\det \| g_{\lambda\mu} \| = 8 (l^0_0)^2 | \det \| \bv_{\alpha} \cdot \bv_{\beta} \| |^{1/2}$ (at $l^0_{\alpha} = 0, \alpha = 1, 2, 3$ assumed) means the power of $| \det \| \bv_{\alpha} \cdot \bv_{\beta} \| |$ as well.

Note also that the factors $|\det \| \bv_{\alpha} \cdot \bv_{\beta} \| |^{3/2}$ also arise on three-dimensional faces when further reducing (in the most symmetrical way) the area tensor part of the measure to the physical hypersurface in the configuration superspace of area tensors. The matter is that the number of triangles is considerably larger than the number of edges. It is seen for the simplest periodic simplicial decomposition when the spacetime is divided into 4-cubes and each 4-cube is divided into 24 4-simplices\cite{RocWil}. Such structure contains 15 edges and 50 triangles per vertex. Inside a 4-simplex its 10 edge lengths locally can be expressed in terms of 10 triangle areas, but without taking into account relations between the different areas the different 4-simplices may not coincide on their common faces. When expressing area tensors in terms of edge vectors, superfluous integrations inevitably get transformed into scale factor of positive dimensionality in $v$ (once reducing the measure is performed in scale-invariant way, by inserting scale-invariant $\delta$-factor). Relations between the different areas are very complicated and nonlocal. Formally more simple way to take them into account in the measure is to extend the set of variables $\vstw$ simultaneously imposing simple conditions. Namely, insert for each triangle $\stw$ a full set of its area tensors defined in the local frames of all the 4-simplices $v_{\stw |\sfo }$ containing this triangle, $\sfo \supset \stw$, and viewed as independent variables. Each area tensor denoted above as $\vstw$ on which the action depends is defined in certain 4-simplex $\sfo = \sfo (\stw )$, $\vstw \equiv v_{\stw | \sfo (\stw )}$. The action does not depend on $v_{\stw | \sfo }, \sfo \neq \sfo (\stw )$. So we can identically insert integrations over $L^{-6}\d^6 v_{\stw | \sfo }, \sfo \neq \sfo (\stw )$ in the measure as well, restricting each component $v^{ab}_{\stw | \sfo}$ to vary in the region $[-L/2, L/2], L \to \infty$. After such extension of the set of variables the condition of the coincidence of six pairs of edge lengths of the pair of the 4-simplices on their common 3-face (tetrahedron) is written simply as six {\it bilinear} conditions for three area vectors defined inside these 4-simplices, $\bv_{\alpha}$ and $\bv^{\prime}_{\alpha}$: $\bv_{\alpha} \cdot \bv_{\beta} = \bv^{\prime}_{\alpha} \cdot \bv^{\prime}_{\beta}$. Take the integration element (\ref{d3v-det}) and the analogous one with $\bv^{\prime}_{\alpha}$, $|\det \| \bv^{\prime}_{\alpha} \cdot \bv^{\prime}_{\beta} \| |^{3/2} \prod_{\alpha} \d^3 \bv^{\prime}_{\alpha}$. Insert the most symmetrical $\delta$-factor $\desi (\bv_{\alpha} \cdot \bv_{\beta} - \bv^{\prime}_{\alpha} \cdot \bv^{\prime}_{\beta}) (\det \| \bv_{\alpha} \cdot \bv_{\beta} \|)^2$. The multiplier of $\delta$-function is singled out by invariance of the overall $\delta$-factor w. r. t. an arbitrary redefinition of the tetrahedron edges $\bl_{\alpha} \to B_{\alpha}{}^{\beta} \bl_{\beta}$, $\bl^{\prime}_{\alpha} \to B_{\alpha}{}^{\beta} \bl^{\prime}_{\beta}$ with the {\it same} $B_{\alpha}{}^{\beta}$ (the viewpoint is undertaken that the primed and unprimed edges are the {\it same} in the world coordinates, while tangential (affine) metric on the common face in the 4-simplices $g_{\alpha\beta} = \bl_{\alpha} \cdot \bl_{\beta}, g^{\prime}_{\alpha\beta} = \bl^{\prime}_{\alpha} \cdot \bl^{\prime}_{\beta}$ undergoes discontinuity\cite{Kha4,Kha4-prime} $\Delta g_{\alpha\beta} = g_{\alpha\beta} - g^{\prime}_{\alpha\beta}$). The product of (\ref{d3v-det}) and $|\det \| \bv^{\prime}_{\alpha} \cdot \bv^{\prime}_{\beta} \| |^{3/2} \prod_{\alpha} \d^3 \bv^{\prime}_{\alpha}$ transforms into
\begin{equation}\label{d3v-det9}
|\det \| \bv_{\alpha} \cdot \bv_{\beta} \| |^{9/2} \prod_{\alpha} \d^3 \bv_{\alpha}
\end{equation}

\noindent (up to integration over 3 rotational degrees of freedom of the triad $\bv^{\prime}_{\alpha}$). According to the above said, the relatively large dimensionality in $v$ of this factor is eventually due to the approach adopted proceeding through the extended configuration superspace of (the large number of the components of) independent area tensors where the latter are true canonical variables. The integration over connections of the present paper relies on this concept as well, but indirectly, as possibility to analytically continue the result to independent area tensors.

Turning to the link suppression, there is possibility that the triangles with small area but the two sides of large length (spikes) might be dominating as is the case in the two-dimensional gravity\cite{Savvidy}. In our case, as considered in subsection \ref{GenCase}, exponential suppression takes place over $\bv_{\alpha}, \alpha = 1, 2, 3$ (area tensors $v_{0\alpha}$ being considered as fixed arguments). This means suppression over $\bl_{\alpha}$s as well, since small $\bv_{\alpha}$s mean small $\bl_{\alpha}$s, with exception of the case when $|\bl_1 \times \bl_2 \cdot \bl_3| = 2 \sqrt{2} |\det \| \bv_{\alpha} \cdot \bv_{\beta} \| |^{1/4} \to 0$ (tetrahedron is flattened). The latter case is suppressed by occurrence of the factor $|\det \| \bv_{\alpha} \cdot \bv_{\beta} \| |^{9/2}$ in the integration element (\ref{d3v-det9}). With this factor, the powers of edge vector components $\bl_{\alpha} = \sqrt{2} \epsilon_{\alpha}{}^{\beta \gamma} \bv_{\beta} \times \bv_{\gamma} |\det \| \bv_{\alpha} \cdot \bv_{\beta} \| |^{-1/4}$ have convergent at $\det \| \bv_{\alpha} \cdot \bv_{\beta} \| \to 0$ expectations $< l^n >$ at $n < 20$. Were we in two dimensions, we would have two edge vectors of the triangle $\bl_1, \bl_2$ and only one area vector $\bl_1 \times \bl_2$ (more exactly, scalar). Suppression of the latter would not be sufficient to guarantee suppression of $\bl_1, \bl_2$.

\subsection{Local maxima}

Appealing feature of the definition of integrals through moments adopted is that when computing any moment we only deal with values of finite number of derivatives of $g(x)$ at $x = 0$, i. e. with local properties of '$\arcsin$' function at $x = 0$ at each step. This is important since thus far path integral formalism has been checked in physics only on perturbative level. Nevertheless, knowing full infinite set of moments somehow takes into account analytical features of this function. In particular, we observe poles of $\N_0 (\rv, \rv^*)$ (\ref{Nvv}) at $\rv^2\!$ $\!= 4n^2(1 + i/\gamma)^{-2}, n\!$ $\!= 1, 2, ...$. Corresponding dependence on $\rv$ looks as result of some summation in the path integral over branches of the '$\arcsin$' function, as if we had substituted $\arcsin \to \arcsin + 2\pi n$ in the exponential and summed over $n$ for each of the two '$\arcsin$' functions. This would just result in the hyperbolic or trigonometric function in the denominator of (\ref{Nvv}). Thus, knowing full set of moments reproduces features related to the global analytical properties of the '$\arcsin$' function.

In the physical region $\Im \rv^2 = 0$ proximity of the poles at $\rv^2 = 4n^2(1 + i/\gamma)^{-2}$ shows up especially at $\gamma \ll 1$ or at $\gamma \gg 1$ as local maxima of $\N_0 (\rv, \rv^*)$ approximately at $\rv^2 = -4 \gamma^2 n^2$ for $\gamma \ll 1$, $n \ll \gamma^{-1}$ or at $\rv^2 = 4 n^2$ for $\gamma \gg 1$, $n \ll \gamma$. It is interesting to compare location of these maxima with eigenvalues of the operator of $\rv$ (in the continuous time limit formalism). In Ref. \cite{Kha1} we have considered kinetic term ${\rm Tr} (v \OmT \dotOm )$ (3-dimensional version of this one was deduced by Waelbroeck\cite{Wael} in lattice gravity from symmetry considerations) in the canonical form of Regge action. Now we only have to rewrite this for nonzero $\gamma^{-1}$,
\begin{eqnarray}
\hspace{-5mm}v \! \circ \! \Omega^{\rm
T}\dot{\Omega} \Rightarrow \left (1 + \frac{i}{\gamma} \right ) \frac{\pbv \! * \! \,^{+}\!(\Omega^{\rm T}\dot{\Omega})}{2} + \left (1 - \frac{i}{\gamma} \right ) \frac{\mbv \! * \! \,^{-}\!(\Omega^{\rm T}\dot{\Omega})}{2} =  \left (v - \frac{*v}{ \gamma} \right ) \! \circ \! \Omega^{\rm
T}\dot{\Omega}
\end{eqnarray}

\noindent for combination of representations $\pS$, $\!\! \mS \,\,$ (\ref{S+S}). Consider the following degrees of freedom in $\Omega$,
\begin{equation}
\Omega \Longrightarrow \Omega \exp \left (\alpha \frac{ v }{ |\rv | } + \beta \frac{* v }{ |\rv | } \right )
\end{equation}

\noindent (rotation {\it around} the triangle parameterized by $\alpha$ and {\it in} the triangle parameterized by $\beta$ preserving the area tensor). Since $v * v = 0$ for physical $v$, contribution of $\alpha, \beta$ to the kinetic term is
\begin{equation}
\frac{\rv^2 }{ |\rv | } \left ( \dot{\alpha} - \frac{\dot{\beta} }{ \gamma}\right ).
\end{equation}

\noindent For the spacelike area $\rv^2 = -|\rv^2|$. Then $\beta$ is Euclidean angle, and requiring the canonical commutation relations and periodicity at $\beta \to \beta + 2\pi$ gives eigenvalues of the conjugate variable $|\rv |/ \gamma$: $|\rv | = \gamma n$. For the timelike area $\rv^2 = |\rv^2|$. Then $\alpha$ is Euclidean angle, and requiring the canonical commutation relations and periodicity at $\alpha \to \alpha + 2\pi$ gives eigenvalues of the conjugate variable $|\rv |$: $|\rv | = n$.

It is seen that the $\rv$ ensuring local maximum in the probability distribution of areas turns out to be approximately the eigenvalue of the operator of $\rv$ within operator approach to the continuous time simplicial gravity. This should not seem surprising since both effects are caused by similar reasons, multivalent nature of '$\arcsin$' function and periodicity over (Euclidean) angles.

\section{Conclusion}

To resume, despite exponential growth of the Haar measure, integration over connection in the path integral in the discrete simplicial gravity in Minkowsky spacetime can be well defined. The result shows that contribution of large areas is suppressed exponentially. This provides also suppression of large edge lengths which is important for that the minisuperspace system described by elementary lengths/areas be self-consistent. The set of local maxima in the probability distribution for area has relation to the set of eigenvalues of the operator of area in the continuous time theory.

\section*{Acknowledgments}

The present work was supported in part by the Russian Foundation for Basic Research
through Grants No. 08-02-00960-a and No. 09-01-00142-a.


\end{document}